\documentclass[letter]{aa}

\usepackage{txfonts}
\usepackage{hyperref}   
\usepackage{longtable}
\usepackage{pifont}
\usepackage{adjustbox}
\usepackage{threeparttable}
\usepackage{graphicx,url,twoopt,natbib}
\usepackage{gensymb}
\usepackage{rotating}
\usepackage{soul}
\usepackage{adjustbox}
\usepackage{threeparttable}
\usepackage{graphicx,url,twoopt,natbib}       
\usepackage{hyperref}   
\usepackage{longtable}
\usepackage{pifont}
\usepackage{xcolor}

\hypersetup{
  colorlinks=true,   
  urlcolor=blue,     
  linkcolor=blue,    
  citecolor=blue}

\usepackage{txfonts}
\usepackage{amsfonts}
\usepackage{caption}
\usepackage{subcaption} 
\usepackage{array}

\newcommand{\kms}{\mbox{km\,s$^{-1}$}}
\newcommand{\mum}{\mbox{$\mu$m}}

\def\snu#1{\ifmmode {S_\nu\,\propto\,\nu^{#1}}
          \else \hbox{$S_\nu$\,$\propto$\,$\nu^{#1}$}\fi}
\def\cm#1{\ifmmode {\,{\rm cm^{-#1}}}                  
        \else \hbox{$\,${\rm cm$^{\rm -#1}$}}\fi}
\def\raw {\ifmmode\rightarrow\else$\rightarrow$\fi}
\def\ex#1{\ifmmode {\cdot 10^{#1}}         
        \else \hbox{{$\times 10^{\rm #1}$}}\fi}

\renewcommand{\descriptionlabel}[1]{\hspace{\labelsep}\textbf{#1}}

\begin{document} 

   \title{MATISSE unveils a dusty disc in the interior of EP Aquarii}
   \subtitle{First evidence of a misaligned hierarchical triple system with a central AGB star?}

   \author{J. Alonso-Hernández \inst{1, 2} \and C. Paladini \inst{3} \and C. Sánchez Contreras \inst{1} \and  J. Drevon \inst{3} \and J. P. Fonfría \inst{4} \and L. Planquart \inst{5} \and T. Khouri \inst{5} \and L. Siess \inst{6} \and A. Jorissen \inst{6} \and B. Aringer \inst{7} \and F. Baron \inst{8} \and E. De Beck \inst{5} \and S. Höfner\inst{9} \and J. Hron \inst{7} \and F. Lykou \inst{10} \and F. Millour \inst{11} \and M. Montarg\`es\inst{12} \and N. Nardetto \inst{11} \and K. Ohnaka \inst{13}\and H. Olofsson \inst{5} \and V. Răstău \inst{7} \and J. Sanchez-Bermudez \inst{14} \and L. Velilla-Prieto \inst{4} \and W. Vlemmings \inst{5} \and M. Wittkowski \inst{15} \and G. Weigelt \inst{16} \and W. C. Danchi\inst{17} \and F. Kerschbaum\inst{7} \and B. Lopez \inst{11}} 

   \institute{Centro de Astrobiología, CSIC-INTA, Camino bajo del Castillo s/n, 28692 Villanueva de la Cañada, Madrid, Spain \and Escuela de Doctorado UAM, Centro de Estudios de Posgrado, Universidad Autónoma de Madrid, 28049 Madrid, Spain \\ 
\email{jaime.alonsoh@estudiante.uam.es}  \and European Southern Observatory, Alonso de Córdova 3107, Vitacura, Santiago, Chile \and Dept. de Astrofísica Molecular, Instituto de Física Fundamental, IFF-CSIC, C/ Serrano 121, 28006 Madrid, Spain \and Department of Physics and Astronomy, Chalmers University of Technology, 412 96 Gothenburg, Sweden \and Institute of Astronomy and Astrophysics, Université libre de Bruxelles, CP 226, Boulevard du Triomphe, 1050 Brussels, Belgium \and Department of Astrophysics, University of Vienna, Türkenschanzstrasse 17, 1180 Vienna, Austria \and Center for High Angular Resolution Astronomy, Georgia State University, CA, 91023, USA \and Theoretical Astrophysics, Department of Physics and Astronomy, Uppsala University, Box 524, 751 20 Uppsala, Sweden \and Konkoly Observatory, HUN-REN CSFK, Konkoly Thege Miklos ut 15-17, 1121 Budapest, Hungary \and Université Côte d'Azur, Observatoire de la Côte d'Azur, CNRS, Laboratoire Lagrange, 06304 Nice, France \and LIRA, Observatoire de Paris, Université PSL, Sorbonne Université, CNRS, 92195 Meudon, France \and Instituto de Astrofísica, Universidad Andrés Bello, Fernández Concha 700, Las Condes, Santiago, Chile \and  Universidad Nacional Autónoma de México. Instituto de Astronomía. A.P. 70-264, 04510 Mexico city, Mexico \and European Southern Observatory, Karl-Schwarzschild Str. 2, 85748 Garching, Germany \and Max Planck Institute for Radio Astronomy, Auf dem H\"ugel 69, 53121 Bonn, Germany \and NASA Goddard Space Flight Center, Astrophysics Division, Greenbelt, MD, 20771, USA
\\
}
\date{Received 6 August 2026 / Accepted 6 October 2026}
  \abstract   {}
   {We have mapped the dust and molecular gas emission distributions in the vicinity (a few stellar radii) of the asymptotic giant branch (AGB) star EP\,Aquarii, with a previously reported stellar companion at $\sim$50\,au, to search for binary-induced asymmetries.}
   {We present an analysis of new high-angular-resolution ($\gtrsim$4\,mas) images obtained with MATISSE (in the wavelength range $\sim$3--8\,\mum) as well as archival high-angular-resolution ($\gtrsim$20\,mas) ALMA $\rm^{12}$CO\,(c$J$=2-1) line and continuum emission observations ($\sim$230\,GHz), to study the circumstellar morphology and kinematics of the molecular gas on these spatial scales.}
   {The MATISSE images display a bright central source and two slightly extended emission features ("blobs") located on opposite sides of the AGB star (at radial distances of $\sim$4--10\,au). These structures are consistent with previous mid-infrared interferometric observations, suggesting stability over decades. The central source is elongated along the direction defined by the blobs, in both the MATISSE and ALMA continuum images. The ALMA $\rm^{12}$CO\,($J$=2-1) maps display an asymmetric inner wind with a velocity gradient.}
   {The observations are consistent with an edge-on dusty equatorial overdensity (disc-like), hinting the presence of a close ($\lesssim$4\,au) companion. The inferred symmetry axis is almost perpendicular to the CO velocity gradient and the orbital plane proposed for the distant ($\sim$50\,au) companion. EP\,Aqr may represent the first misaligned hierarchical triple system with a central AGB star.}


   \maketitle
   \nolinenumbers

\section{Introduction}

Low- and intermediate-mass stars (i.e. with initial masses $\sim$1--8\,$M_{\odot}$) evolve through the Asymptotic Giant Branch (AGB) undergoing intense mass-loss, leading to the formation of dense and dusty circumstellar envelopes (CSEs) surrounding the stars \citep{Hofner_2018}. Recent high-angular-resolution observations have unveiled the presence of faint asymmetrical structures in the CSEs such as spirals and arcs \citep[see e.g.][]{Decin_2020, Danilovich_2025}, which indicate the presence of stellar companions that shape the circumstellar material resulting in these structures \citep[see][]{Mastrodemos_1999, Mohamed_2007}.

This study focuses on the oxygen-rich (O-rich) AGB star EP\,Aquarii (EP\,Aqr), a semi-regular pulsator with multiple  periodicity components in the range 80--130\,days \citep{Tabur_2009}, a systemic velocity ($V_{\mathrm{sys}}$) of -34\,\kms\ in the local standard of rest (LSR) frame, and an estimated mass-loss rate of $\sim$1.2$\times 10^{-7}\,$\,$M_{\odot}\rm{yr}^{-1}$ \citep[see e.g.][]{Nhung_2015}. Gaia parallax measurements give a distance of 130$\pm$40\,pc (i.e. 1\,au corresponds to $\simeq$7.7\,mas) for EP\,Aqr \citep{Gaia, Andriantsaralaza_2022}.

EP\,Aqr is the archetype of a subclass of AGB stars with unusual CO sub-mm emission line profiles, composed of a broad ($>$10\,$\rm km\,s^{-1}$) parabolic component and a narrow ($\leq$5\,$\rm km\,s^{-1}$) central peak, which were first interpreted as the result of two consecutive mass-loss stages \cite[see][]{Knapp_1998}. Alternatively, \cite{Kerschbaum_1999} suggested the presence of bipolar outflows or rotating discs. Recent Atacama Large Millimeter/submillimeter Array (ALMA) observations of EP\,Aqr revealed a more complex reality. \cite{Homan_2018} found a large spiral (with a diameter of $\sim$10$^{\prime\prime}$) oriented almost face-on (with an inclination angle between the orbital axis and the plane of the sky of $\sim$4--18\degree) inscribed in its CSE, presumably shaped by a stellar companion located at a radial distance of $\sim$0.4$^{\prime\prime}$ ($\sim$50\,au) from the AGB star, that is revealed by a prominent off-centre void in the SiO emission. A complex velocity gradient traced by CO and SO$_{2}$ lines in the innermost ($\sim$0.1--0.4$^{\prime\prime}$) region of the CSE was interpreted as a rotating structure \citep[][]{Homan_2018, Homan_2020}. The presence of a nearly pole-on bipolar outflow along a position angle (PA) $\sim$160\degree\,in EP\,Aqr was also suggested \citep[see][]{Hoai_2019, Tuan-Anh_2019, Nhung_2024}

In this letter, we present new high-angular-resolution observations of EP\,Aqr carried out with the Multi AperTure mid-Infrared SpectroScopic Experiment \citep[MATISSE,][]{Lopez_2022} instrument installed on the Very Large Telescope Interferometer (VLTI). We focus on the description and interpretation of multi-wavelength images in the $\sim$3--8\,\mum\ range, which reveal the emission distribution of dust and some molecular species in the innermost 100 mas ($\sim$13\,au) environment. We complement these images with archival continuum and ALMA $\rm^{12}$CO\,($J$=2-1) observations to achieve a more comprehensive description. 

\section{Observations}

\subsection{MATISSE}

Observations were carried out in August-September 2019 in standalone mode as part of the BIN-AGB Large Program (ID 108.22E9, PI: C. Paladini), which aims to image the innermost environments of a sample of nearby AGB stars to study the impact of stellar companions on dust formation and stellar winds. EP\,Aqr was observed at low spectral resolution (R$\simeq$30), in the L (2.8--4.2), M (4.5--5.0\,\mum) and N (8.0--13.0\,\mum) bands, with the four auxiliary telescopes (ATs) in four different configurations: small (A0-B2-D0-C1), medium (K0-G2-D0-J3), and large (A0-G1-J2-J3 and A0-G1-J2-K0), resulting in an angular resolution down to 4\,mas. 

The observations cover a timespan of 14\,days, which is $\sim$12\% of the main pulsation period of EP\,Aqr \citep[$\sim$118\,days, see][]{Tabur_2009}. Therefore, we combined all the available good quality data since it is not expected that significant morphological changes occurred within this time interval. A summary of the observations and the calibrators is available in Appendix~\ref{OBS_log}. 

We performed a standard data reduction using the Python package \texttt{mat\_tools}\footnote{\url{https://github.com/Matisse-Consortium/tools}} developed by the MATISSE consortium, which acts as a wrapper around the standard ESOREX data-reduction software (version 3.13.6) for ESO/MATISSE. For the image reconstruction, we used the Multi-aperture Image Reconstruction Algorithm \citep[MiRA,][]{Thiebaut_2008}, which was specifically developed for optical interferometric image restoration.

We performed the image reconstruction following the two-step methodology described in \cite{Drevon_2026}. As a first step, we used the \texttt{PYRA}\footnote{\url{https://github.com/jdrevon/PYRA}} (Python for MiRA) tool to generate a sample of 1000 reconstructed images with 500 iterations to sample the parameter space (e.g. field of view, pixel scale, hyper-parameter), applying a quadratic compactness regulariser. In a second step, we used \texttt{MYTHRA}\footnote{\url{https://github.com/jdrevon/MYTHRA}} (Mean Astrophysical Images with \texttt{PYRA}), which generates a mean image with the optimal value of the hyper-parameter and its respective error map.

In the L-band, image reconstruction was restricted to wavelengths longer than 3.1\,\mum\ due to the limited signal-to-noise ratio. For the N-band, it was restricted to the spectral range $\sim$8.0-8.5\,\mum\ due to the lack of high spatial frequency coverage at longer wavelengths (see Appendix.~\ref{observables}).

\subsection{ALMA archival observations}

EP\,Aqr was observed with the ALMA 12m array in June 2019 under the project 2018.1.00750.S (PI: W. Homan). These observations were carried out with the most extended configurations (C-9 and C-10) in Band 6 (1.3\,mm) with a total of four spectral windows (with central frequencies at 217.128, 220.422, 230.562 and 235.176 GHz). A detailed description of the observing setup was presented in \cite{Homan_2020}.

Calibrated measurement sets were used to produce visibility ($u$,$v$) tables, exported to \texttt{Gildas}\footnote{\url{https://www.iram.fr/IRAMFR/GILDAS/}} format for imaging, cleaning and restoration. We used a line-subtracted continuum image to perform self-calibration, improving the signal-to-noise by a factor $\sim$2. We also obtained $\rm^{12}$CO\,($J$=2-1) line data after subtracting the continuum. The ALMA images presented in this letter were restored using the H\"ogbom CLEAN method \citep{Hogbom_1974}. We used robust parameter of 1.0 and 0.1 (natural and uniform weighting, hereafter) for continuum and $\rm^{12}$CO\,($J$=2-1) line maps, respectively, achieving beam sizes of $\sim$30\,mas$\times$39\,mas and $\sim$20\,mas$\times$22\,mas in each case.

\section{Results: A highly-asymmetric environment}

We performed image reconstruction of the MATISSE data at six specific wavelengths, which were selected to illustrate the pseudo-continuum (P-C) emission and the effects of the various vibrational molecular bands in this wavelength range (see Appendix.~\ref{images}): (i) 3.12\,\mum\ for $\rm H_{2}O$ ($\Delta v_{1,3}$=1); (ii) 3.92\,\mum\ for L-band P-C; (iii) 4.06\,\mum\ for SiO ($\Delta v$=2); (iv) 4.59\,\mum\ for CO ($\Delta v=1$); (v) 4.96\,\mum\ for M-band P-C; and (vi) 8.01\,\mum\ for N-band P-C. The images that cover the molecular bands are not continuum-subtracted due to the low spectral resolution. The estimated root mean squared (rms) noises of the images are of 1\% in the L- and M-bands images, and 0.6\% in the N-band image.

\begin{figure}[h!]
    \centering
    \includegraphics[width=0.93\linewidth]{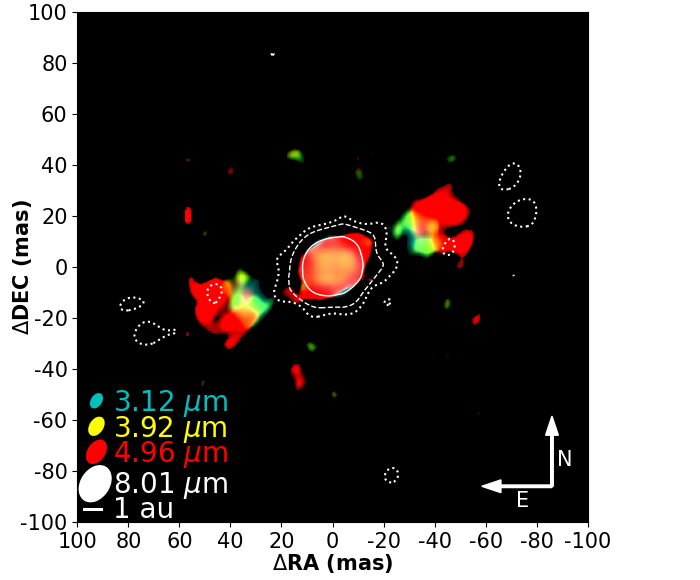}
    \caption{Colour-composite image combining four MATISSE images. Cyan: 3.12\,\mum, Yellow: 3.92\,\mum, Red: 4.96\,\mum. White contours (solid, dashed, and dotted) represent the 3, 10, and 50 rms levels of the 8.01\,\mum\ image. Beam sizes and linear scale (1\,au) are shown at the bottom-left as coloured ellipses and a white line. PA increases from north to east.}
    \label{fig:RGB_image}
\end{figure}

Two main morphological components can be readily identified in the reconstructed images (see Fig.~\ref{fig:RGB_image}): (i) a bright and partially resolved central source that is slightly elongated (along  PA$\sim$130$\degree$); (ii) two slightly extended emission features (hereafter referred to as "blobs") located on opposite sides of the central source along a similar axis (covering PA$=$115$\degree$$\pm$15$\degree$).

Given the sparse ($u$,$v$)-plane coverage, we fitted the observed visibilities by a parametric model composed of a central disc with an elliptical component plus an elongated ring representing the blobs, showing they are not artefacts. (see Appendix.~\ref{UV_fitting} and Figs.~\ref{fig:model_image} and~\ref{fig:model_observales}). We also checked that the presence and locations of the blobs are consistent in sub-samples of the data.

The central source contains the emission of the AGB star. In the P–C images, its apparent size increases with wavelength, consistent with the behaviour inferred from the interferometric observables (see Appendix~\ref{observables}), with a (3 rms level) radius of 9.4$\pm$0.5\,mas (i.e.\,$\sim$1.2\,au) at 3.92\,\mum, 12.9$\pm$2.1\,mas (i.e.\,$\sim$1.7\,au) at 4.96\,\mum, and 19.0$\pm$2.0\,mas (i.e.\,$\sim$2.5\,au) at 8.01\,\mum. This trend is expected because the outer atmospheric layers contribute more at longer wavelengths. The central source is slightly larger in the wavelengths that fall within the different molecular bands (H$_{2}$O, SiO, and CO) than in their nearest P-C images, which may be partially due to emission from the extended atmosphere \citep[similar to that shown in][from CO rotational lines]{Khouri_2024}. We note that it is elongated in the SiO and CO images, whereas it has a round shape in the H$_{2}$O image (see Fig.\ref{MATISSE_images}).

Regarding the pair of blobs, they shift radially with wavelength, covering radial distances from the centre of the star of $\sim$30--40\,mas (i.e.\,$\sim$4--5\,au) at 3.12\,\mum\ and $\sim$45--80\,mas (i.e.\,$\sim$6--10\,au) at 8.01\,\mum\ (see Fig.\ref{fig:RGB_image}). The blobs are slightly brighter (relative to the star) in the molecular bands than in their nearest pseudo-continuum (Fig.~\ref{MATISSE_images}), but it is not possible to analyse the molecular emission due to the low spectral resolution and the uncertain continuum contribution.

The ALMA continuum image at 1.3\,mm (Fig.~\ref{fig:MATISSE+cont}) shows a partially resolved central source that appears to be elongated in the same direction as in the MATISSE images. However, there is no clear 1.3\,mm continuum counterpart to the blobs, constraining their spectral index to $>$2 (see Appendix~\ref{ALMA_continuum}).

Figure.~\ref{fig:ALMA_comparison_CO} shows the ALMA moment 0 map of the $\rm^{12}$CO\,($J$=2-1) line, displaying a bright extended atmosphere surrounding the AGB star. Similar morphologies were found in three O-rich AGB stars: R\,Leo\,\citep{Fonfria_2019}, W\,Hya \citep{Khouri_2019, Ohnaka_2025}, and R\,Dor \citep{Khouri_2019, Khouri_2024}. We highlight that the extended atmosphere is asymmetric and exhibits two regions of maximum emission aligned with the blobs.

Figure~\ref{fig:ALMA-velo} shows the ALMA moment 1 map of the $\rm^{12}$CO\,($J$=2-1) line to illustrate the kinematics of the circumstellar gas on these spatial scales. There is a linear velocity gradient inside the inner wind region ($\sim$100\,mas), consistent with that inferred from the lower angular resolution maps by \cite{Homan_2018, Homan_2020}, whose direction is almost perpendicular to the blobs. 

\section{Discussion: physical interpretation}

The most striking feature of the MATISSE images is the pair of blobs, which is not expected to be around an isolated AGB star. The fact that their relative flux with respect to the star is similar in the different images (including P-C and molecular bands), their radial shift with wavelength, and the lack of a counterpart in the ALMA 1.3\,mm continuum indicates that they are most likely produced by dust thermal emission and, perhaps, some contribution from molecular bands, in which the blobs are brighter.

EP\,Aqr was previously observed in the N-band with the MID-infrared Interferometric instrument \citep[MIDI, see][]{Leinert_2003} in October-December 2012 (i.e.\,$\sim$10\,years before our observations). Even though it was not possible to carry out image reconstruction over this dataset, the geometric modelling presents a solution with a circular uniform disc (diameter of $\sim$11.3\,mas) and an elliptical Gaussian (full width at half maximum of $\sim$41-82\,mas) whose major axis is oriented along PA$\sim$116\degree\, \citep[see][]{Klotz_PhD}. This solution agrees in shape and size with our reconstructed image at 8.01\,\mum, indicating that the structure is stable (stationary), at least, on timescales of decades. A similar structure was also proposed by \cite{Klotz_2012} for the MIDI observations of the O-rich AGB star SV\,Psc.

We interpret the pair of blobs as the limb-brightened edges of a dusty equatorial overdensity surrounding the star (disc-like) viewed nearly edge-on. Similar morphologies are reported in the sub-mm continuum emission of nearly edge-on circumbinary discs around post-AGB stars \citep[see][]{Bujarrabal_2023} and protoplanetary discs \citep[see][]{Alves_2020}. These discs are stable structures and might show thermal gradients.

\begin{figure}[h!]
    \centering
    \includegraphics[width=0.93\linewidth]{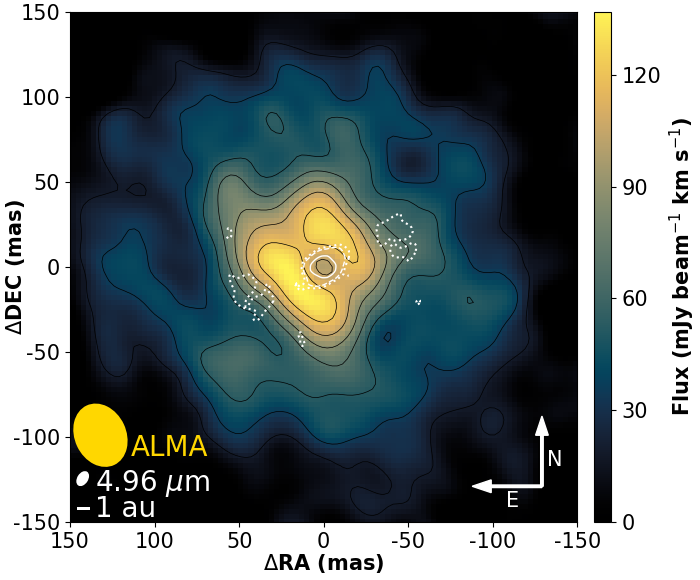}
    \caption{ALMA $\rm^{12}$CO\,($J$=2-1) moment 0 map integrated over a velocity range of $V_{\mathrm sys}$$\pm$30\,\kms. White contours (solid, dashed, and dotted) represent the 3, 10, and 50 rms levels of the 4.96\,\mum\ image. Beam sizes and linear scale are shown as in Fig.~\ref{fig:RGB_image}.}
    \label{fig:ALMA_comparison_CO}
\end{figure}

Even though the ($u$,$v$)-plane coverage is adequate to establish the overall elongation, limitations in sensitivity and dynamic-range may prevent the detection of fainter emission component (similar to that shown for L$_{2}$ Pup by \cite{Lykou_2015}). This interpretation remains consistent with an internal dust-forming region that can be traced with the $\rm^{12}$CO\,($J$=2-1) line (see Fig.~\ref{fig:ALMA_comparison_CO}).

\begin{figure}
    \centering
    \includegraphics[width=0.93\linewidth]{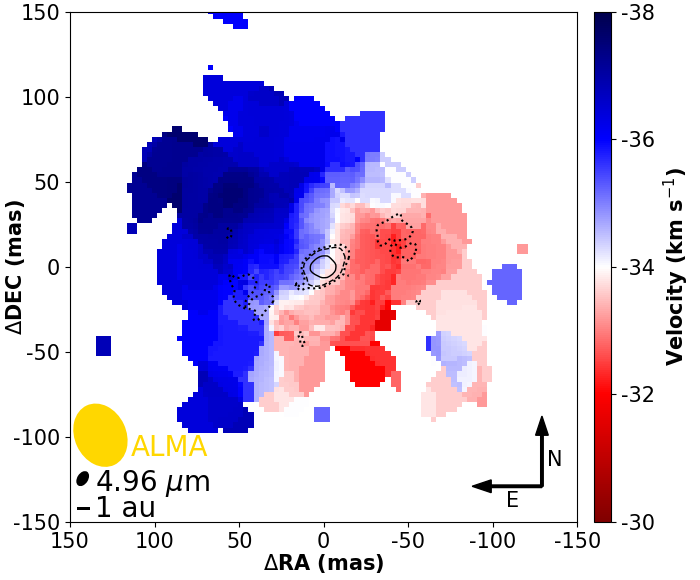}
    \caption{ALMA $\rm^{12}$CO\,($J$=2-1) moment 1 map in the LSR frame around $V_{\mathrm sys}$. Black contours (solid, dashed, and dotted) represent the 3, 10, and 50 rms levels of the 4.96\,\mum\ image. Beam sizes and linear scale are shown as in Fig.~\ref{fig:RGB_image}.}
    \label{fig:ALMA-velo}
\end{figure}

The disc-like structure might be produced by a close companion, which may be also responsible for the elongation of the central source and the extended atmosphere along a common direction (Figs~\ref{fig:RGB_image},  \ref{fig:ALMA_comparison_CO}, and~\ref{fig:MATISSE+cont}). In this scenario, the observed elongation suggest the direction of the orbital plane of the system and the velocity field in the innermost wind (Fig.~\ref{fig:ALMA-velo}) is compatible with a nascent bipolar stream produced by the companion in a direction perpendicular to the orbit.

The proposed companion would be located inside the inner rim of the disc-like structure (i.e.\,$\lesssim$4\,au), consistent with the dust forming region \citep[see][]{Ohnaka_2025}, and its orbit would be contained in the same plane (i.e. an inclination angle of $\sim$70--110\degree, see Appendix.~\ref{UV_fitting}). This orbit would be highly-misaligned with respect to that of the distant ($\sim$50\,au) companion with an inclination angle of 4--18\degree\ reported by \cite{Homan_2018}.

Although misaligned orbits are less common than aligned ones in compact multiple stellar systems, they are commonly reported. In particular, when the ratio between orbital separations is a factor $\gtrsim$10 \cite[see][]{Shariat_2025}. Although orbital misalignments have never been identified before in AGB stars, they are invoked to explain the complex morphologies of certain Planetary Nebulae \citep[see e.g.][]{Soker_2016, De_marco_2022}.

An alternative interpretation is that the blobs trace a bipolar outflow, although these are typically detected in sub-mm molecular lines rather than in mid-infrared thermal emission \citep[as in the case of V\,Hya, see e.g.][]{Sahai_2022, Planquart_2024}. In this scenario, the $\rm^{12}$CO\,($J$=2-1) velocity gradient might trace a rotating structure as suggested by \cite{Homan_2018, Homan_2020}.

Finally, a recent mass ejection episode is rejected due to the
spatial coincidence with the structure suggested by \cite{Klotz_PhD}, which indicates a stability of, at least, a decade. Mass ejection episodes are expected to present larger spatial variability on this timescale \citep[see][]{Freytag_2023, Khouri_2026}.

\section{Conclusions}

We study the AGB star EP\,Aquarii with MATISSE and ALMA observations. We report the detection of two slightly extended mid-infrared emission features located on opposite sides of the AGB star that extend, at least, to $\gtrsim$9\,au. We interpret this structure as the limb-brightened edges of an edge-on dusty equatorial overdensity (disc-like). We found that the AGB star and its extended atmosphere are also elongated in a similar direction. We speculate that this structure is shaped by a close ($\lesssim$4 au) companion whose orbital plane is misaligned with that of a distant ($\sim$50\,au) companion previously reported by \cite{Homan_2018}. 

To date, only a few disc-like structures have been found surrounding AGB stars \citep[see e.g.][]{Kervella_2014, Fonfria_2019, Sanchez-Contreras_2022}. If confirmed, EP\,Aqr may represent the first case of a misaligned hierarchical triple system with a central AGB star, offering new insights for the study of binary-induced structures, dust formation in binary AGB winds. Further observations are needed to confirm these structures and better characterise this system.

\begin{acknowledgements}

We thank the referee, Anita Richards, whose valuable comments helped us to improve the quality of the manuscript. Based on observations collected at the European Southern Observatory under ESO programme 108.22E9.004. This paper makes use of the following ALMA data: ADS/JAO.ALMA\#2018.1.00750.S. ALMA is a partnership of ESO (representing its member states), NSF (USA) and NINS (Japan), together with NRC (Canada), NSTC and ASIAA (Taiwan), and KASI (Republic of Korea), in cooperation with the Republic of Chile. The Joint ALMA Observatory is operated by ESO, AUI/NRAO and NAOJ. This work is part of the I+D+i project PID2023-146056NB-C22 and PID2023-147545NB-I00 funded by Spanish MCIN/AEI/10.13039/501100011033 and by “ERDF A way of making Europe". J.A.H is supported by INTA grant PRE\_MDM\_05 and acknowledges support from the ESO Scientific Visitor Programme.
SH acknowledges funding by the European Research Council (ERC) under the European Union’s Horizon 2020 research and innovation programme (Grant agreement No. 883867, project EXWINGS) and by the Knut and Alice Wallenberg Foundation (grant no. KAW 2023.0314). K.O. acknowledges the support of the Agencia Nacional de Investigación y Desarrollo (ANID) through the FONDECYT Regular grant 1240301. J.S-B. acknowledges the support received by the UNAM DGAPA-PAPIIT project AG-101025 and from the SECIHTI Ciencia de Frontera project CBF-2025-I-3033. L.V-P. acknowledges funding support through grant RYC2023-045648-I.
\end{acknowledgements}

\bibliographystyle{aa} 
\bibliography{main.bib}

\begin{appendix}

\onecolumn

\section{Observation log}\label{OBS_log}

\begin{figure}[h!]
    \centering
    \includegraphics[width=0.93\linewidth]{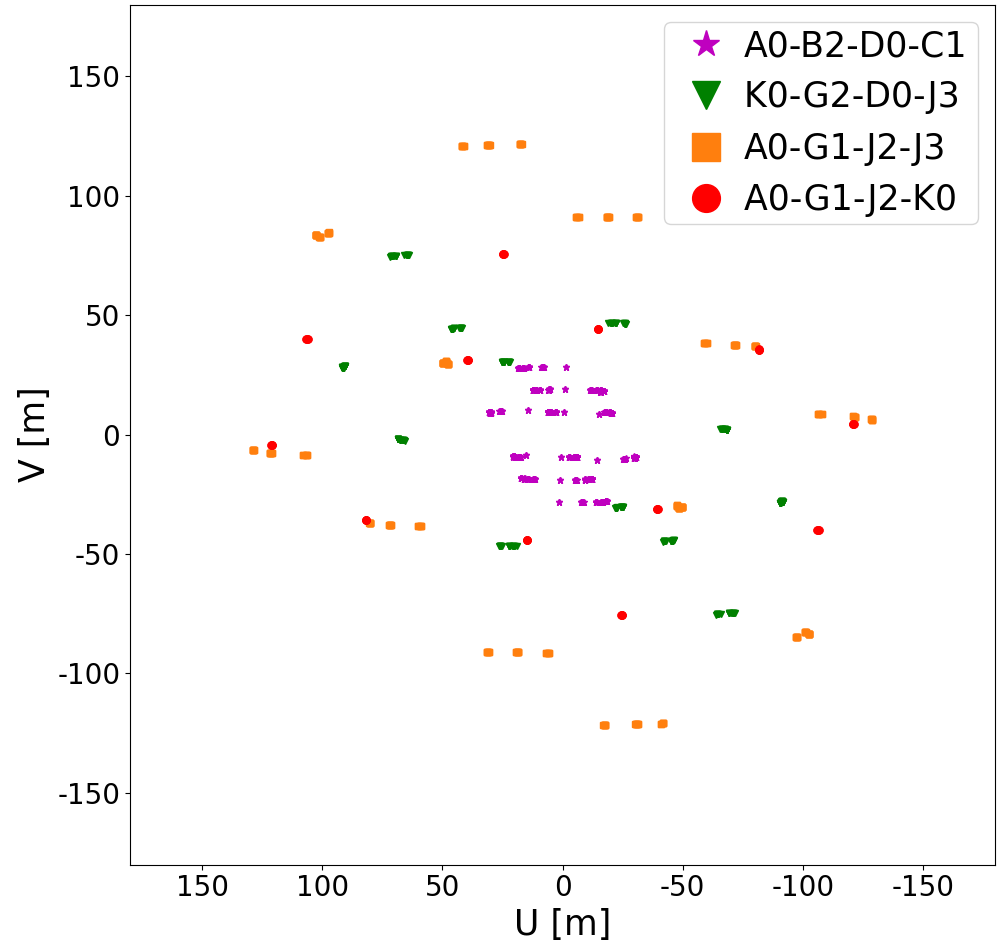}
    \caption{($u$,$v$)-plane coverage of the MATISSE observations in spatial coordinates. The markers colour represents the AT configuration corresponding to each point.}
    \label{fig:uv_coverage}
\end{figure}

We present additional details on the sequence and the atmospheric conditions during the MATISSE observations. Table~\ref{tab:observation_log} shows the observation log sheet, including both the science target and the calibrators. Table~\ref{tab:calibrators} shows the relevant information of the calibrators (3\,Equ, g\,Aqr, k\,Aqr, and lam\,Aqr). Figure~\ref{fig:uv_coverage} shows the ($u$,$v$)-plane coverage obtained for each VLTI configuration.

\begin{table*}[hbt!]

\renewcommand{\arraystretch}{1.1}
\small
\centering

\caption{MATISSE Observation log used for EP\,Aqr} 
\label{tab:observation_log}

\begin{adjustbox}{max width=\textwidth}
\begin{threeparttable}[b]

\begin{tabular}{>{\centering\arraybackslash}p{1.50cm} >{\centering\arraybackslash}p{1.50cm} >{\centering\arraybackslash}p{2.50cm} c c c c c c}
\hline\hline 
Date & Time [UT] & VLTI config. & Object. & seeing [\arcsec] & $\tau_{0}$ & FT & BCD-LM & BCD-N \\   
\hline
2022-08-22 & 03:41:53 & A0-B2-D0-C1   & lam\,Aqr  &  0.76 & 4.49  & 1.00 & 1, 2, 3, 4 & 1, 2$^{*}$, 3, 4 \\
           & 04:05:25 &               & EP\,Aqr   &  1.01 & 3.72  & 1.00 & 1, 2$^{*}$, 3, 4 & 1, 2, 3, 4 \\
           & 05:32:49 &               & EP\,Aqr   &  0.72 & 6.05  & 1.00 & 1, 2, 3, 4 & 1, 2$^{*}$, 3, 4 \\
           & 06:29:38 &               & k\,Aqr    &  1.14 & 2.83  & 1.00 & 1, 2, 3, 4 & 1, 2, 3, 4 \\

2022-08-23 & 02:33:06 & A0-B2-D0-C1   & EP\,Aqr   &  0.61 & 5.32  & 1.00 & 1, 2, 3, 4 & 1, 2, 3, 4 \\
           & 02:58:37 &               & k\,Aqr    &  0.58 & 5.31  & 1.00 & 1, 2, 3, 4 & 1, 2, 3, 4 \\
           & 04:19:06 &               & lam\,Aqr  &  1.78 & 1.74  & 1.00 & 1, 2, 3, 4 & 1, 2, 3, 4 \\
           & 04:43:30 &               & EP\,Aqr   &  1.67 & 1.86  & 1.00 & 1, 2, 3, 4 & 1, 2, 3, 4 \\
           & 05:06:24 &               & lam\,Aqr  &  1.55 & 1.64  & 1.00 & 1, 2, 3, 4 & 1, 2, 3, 4 \\
           & 05:30:45 &               & EP\,Aqr   &  1.78 & 1.58  & 1.00 & 1, 2, 3, 4 & 1$^{*}$, 2, 3, 4 \\

2022-08-25 & 00:13:00 & A0-B2-D0-C1   & k\,Aqr    &  0.59 & 6.70  & 1.00 & 1, 2, 3, 4 & 1, 2, 3, 4 \\
           & 00:35:11 &               & EP\,Aqr   &  0.60 & 7.84  & 1.00 & 1$^{*}$, 2$^{*}$, 3$^{*}$, 4$^{*}$   & 1$^{*}$, 2, 3, 4$^{*}$ \\
           & 04:13:19 &               & k\,Aqr    &  0.43 & 7.86  & 1.00 & 1, 2, 3, 4 & 1, 2, 3, 4 \\
           & 04:37:10 &               & EP\,Aqr   &  0.45 & 8.09  & 1.00 & 1$^{*}$, 2$^{*}$, 3$^{*}$, 4$^{*}$ & 1$^{*}$, 2$^{*}$, 3$^{*}$, 4$^{*}$ \\
           & 05:00:29 &               & k\,Aqr    &  0.49 & 8.74  & 1.00 & 1, 2, 3, 4 & 1, 2, 3, 4 \\
           & 05:22:32 &               & EP\,Aqr   &  0.46 & 9.77  & 1.00 &  1$^{*}$, 2$^{*}$, 3$^{*}$, 4$^{*}$ & 1$^{*}$, 2$^{*}$, 3, 4$^{*}$ \\

2022-08-27 & 03:42:13 & K0-G2-D0-J3   & 3\,Equ    &  0.51 & 9.09  & 1.00 & 1, 2, 3, 4 & -- \\
           & 03:57:04 &               & EP\,Aqr   &  0.41 & 10.08 & 1.00 & 1, 2, 3, 4 & 1$^{*}$, 2, 3, 4 \\
           & 04:19:24 &               & k\,Aqr    &  0.34 & 11.79 & 1.00 & 1, 2, 3, 4 & 1, 2, 3, 4 \\
           & 04:41:57 &               & g\,Aqr    &  0.31 & 11.85 & 1.00 & 1, 2, 3, 4 & -- \\
           & 04:53:22 &               & EP\,Aqr   &  0.35 & 11.01 & 1.00 & 1, 2, 3, 4 & 1$^{*}$,2$^{*}$,3$^{*}$,4$^{*}$ \\
           & 05:15:05 &               & lam\,Aqr  &  0.48 & 9.16  & 1.00 & 1, 2, 3, 4 & 1, 2 ,3, 4 \\

2022-08-29 & 04:15:38 & K0-G2-D0-J3   & g\,Aqr    &  0.35 & 5.93  & 1.00 & 1, 2, 3, 4 & -- \\
           & 04:27:30 &               & EP\,Aqr   &  0.47 & 5.36  & 1.00 & 1$^{*}$, 2, 3, 4$^{*}$ & 1$^{*}$, 2, 3, 4$^{*}$ \\
           & 04:49:24 &               & lam\,Aqr  &  0.42 & 6.68  & 1.00 & 1, 2, 3, 4 & 1, 2, 3, 4 \\

2022-08-31 & 03:48:26 & A0-G1-J2-J3   & 3\,Equ    &  0.44 & 5.95  & 1.00 & 1, 2, 3, 4 & -- \\
           & 03:59:45 &               & EP\,Aqr   &  0.41 & 5.51  & 1.00 & 1, 2, 3, 4 & 1$^{*}$, 2, 3, 4 \\
           & 04:23:32 &               & k\,Aqr    &  0.57 & 4.67  & 1.00 & -- & 1, 2, 3, 4 \\
           & 04:49:15 &               & g\,Aqr    &  0.57 & 4.37  & 1.00 & 1, 2, 3, 4 & -- \\
           & 05:00:12 &               & EP\,Aqr   &  0.62 & 4.40  & 0.99 & 1, 2, 3, 4 & 1, 2, 3, 4 \\
           & 05:22:22 &               & lam\,Aqr  &  0.61 & 4.00  & 1.00 & -- & 1, 2, 3, 4 \\
           & 05:45:13 &               & g\,Aqr    &  0.64 & 4.18  & 1.00 & 1, 2, 3, 4 & -- \\
           & 05:56:28 &               & EP\,Aqr   &  0.76 & 2.90  & 0.98 & 1, 2, 3$^{*}$, 4 & 1, 2, 3$^{*}$, 4 \\
           & 06:19:21 &               & lam\,Aqr  &  0.59 & 3.97  & 1.00 & -- & 1, 2, 3, 4 \\

2022-09-04 & 01:38:07 & A0-G1-J2-K0   & k\,Aqr    &  0.53 & 3.90  & 1.00 & 4 & 1, 2, 3, 4 \\
           & 02:03:37 &               & EP\,Aqr   &  0.57 & 3.33  & 0.99 & 1, 2, 3, 4$^{*}$ & 1, 2, 3, 4$^{*}$ \\
           & 02:28:05 &               & 3\,Equ    &  0.60 & 4.25  & 1.00 & 1, 2, 3, 4 & -- \\
\hline
\end{tabular} 
\begin{tablenotes}
\item \textbf{Notes.} Column~(1): Date of the observation, Col.~(2): Universal time at the observing start of the observing block, Col.~(3): Configuration of the AT stations, Col.~(4): Object observed during the block, Col.~(5): Average seeing during the block, Col.~(6): Average coherence time during the block, Col.~(7): Average fringe tracking ratio, Col.~(8): Beam Commuting Device (BCD) configuration (the numbers 1, 2, 3, and 4 correspond to IN-IN, IN-OUT, OUT-IN, OUT-OUT respectively). $^{*}$ These observations were rejected during data reduction due to poor quality.

\end{tablenotes}

\end{threeparttable}
\end{adjustbox}

\renewcommand{\arraystretch}{1.0}

\end{table*}

\begin{table*}[h!]

\renewcommand{\arraystretch}{1.1}
\small
\centering

\caption{List of the calibrators used for the MATISSE data and their main properties. This information was obtained from the Mid-infrared stellar Diameters and Fluxes compilation Catalogue \citep[MDFC,][]{Cruzalebes_2019}.} 
\label{tab:calibrators}

\begin{adjustbox}{max width=\textwidth}
\begin{threeparttable}[b]

\begin{tabular}{l >{\centering\arraybackslash}p{2.50cm} >{\centering\arraybackslash}p{2.50cm} >{\centering\arraybackslash}p{2.50cm} >{\centering\arraybackslash}p{3.00cm} >{\centering\arraybackslash}p{3.00cm}}
\hline\hline 
Name & Spectral type & $\theta_{\rm UD}$ ($L$) [mas] & $\theta_{\rm UD}$ ($N$) [mas] & $L$-band flux (\#) [Jy] & $N$-band flux (\#) [Jy] \\   
\hline
3\,Equ   & K5III   & 2.375$\pm$0.242 & 2.394$\pm$0.242 &  51$\pm$1   (3) &  9$\pm$1  (6) \\
g\,Aqr   & K3III   & 2.695$\pm$0.234 & 2.717$\pm$0.234 &  75$\pm$7   (3) & 11$\pm$3  (5) \\
k\,Aqr   & M3III   & 7.072$\pm$0.857 & 7.090$\pm$0.857 & 396$\pm$85  (4) & 51$\pm$11 (5) \\
lam\,Aqr & M2.5III & 8.304$\pm$0.910 & 8.331$\pm$0.910 & 537$\pm$321 (4) & 71$\pm$10 (7) \\

\hline
\end{tabular} 
\begin{tablenotes}
\item \textbf{Notes.} Column~(1): Calibrator, Col.~(2): Spectral type, Col.~(3): Uniform Disc Diameter in band $L$, Col.~(4): Uniform Disc Diameter in band $N$, Col.~(5): Median flux value in band $L$ (Number of flux values reported), Col.~(6): Median flux value in band $N$ (Number of flux values reported).
\end{tablenotes}

\end{threeparttable}
\end{adjustbox}

\renewcommand{\arraystretch}{1.0}

\end{table*}

\twocolumn 

\section{MATISSE observables}\label{observables}

\begin{figure*}[h!]
    \centering
    \includegraphics[width=\linewidth]{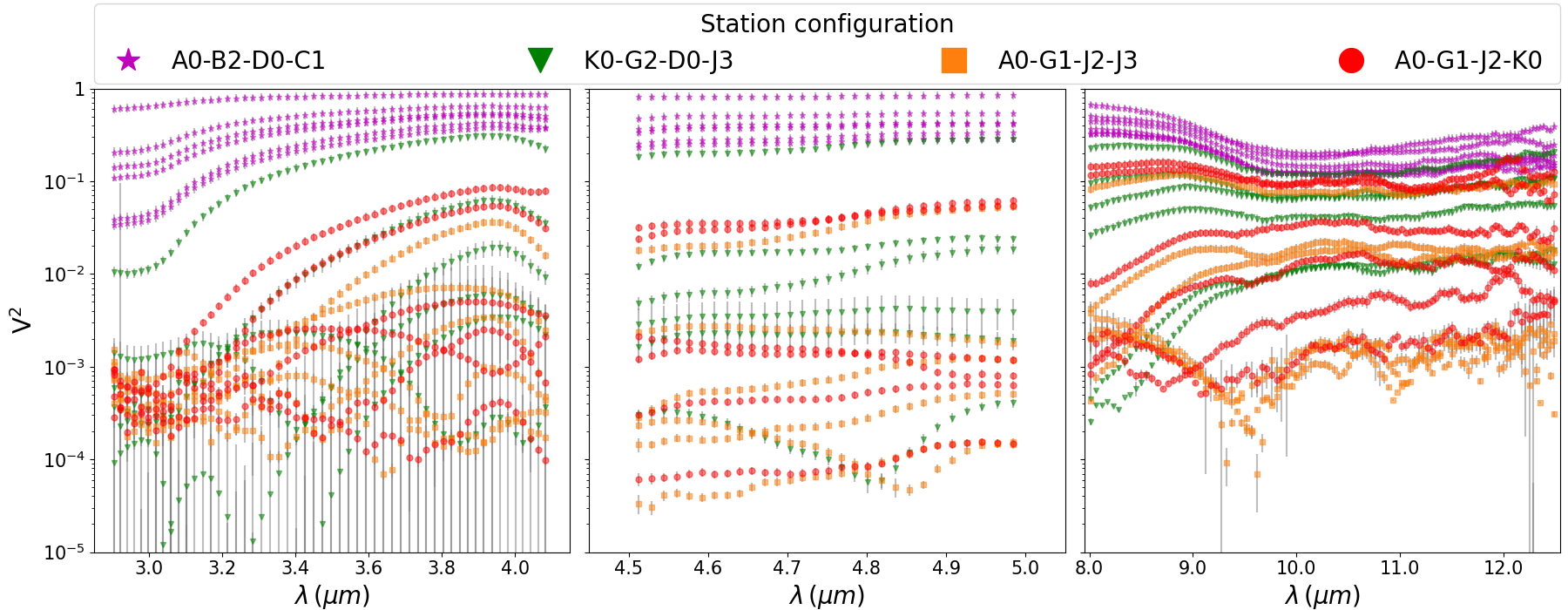}
    \caption{Comparison of the squared visibilities from the MATISSE observations with wavelength. The markers colour represent the AT configuration corresponding to each curve.}
    \label{fig:VIS2}
\end{figure*}

\begin{figure*}[h!]
    \centering
    \includegraphics[width=\linewidth]{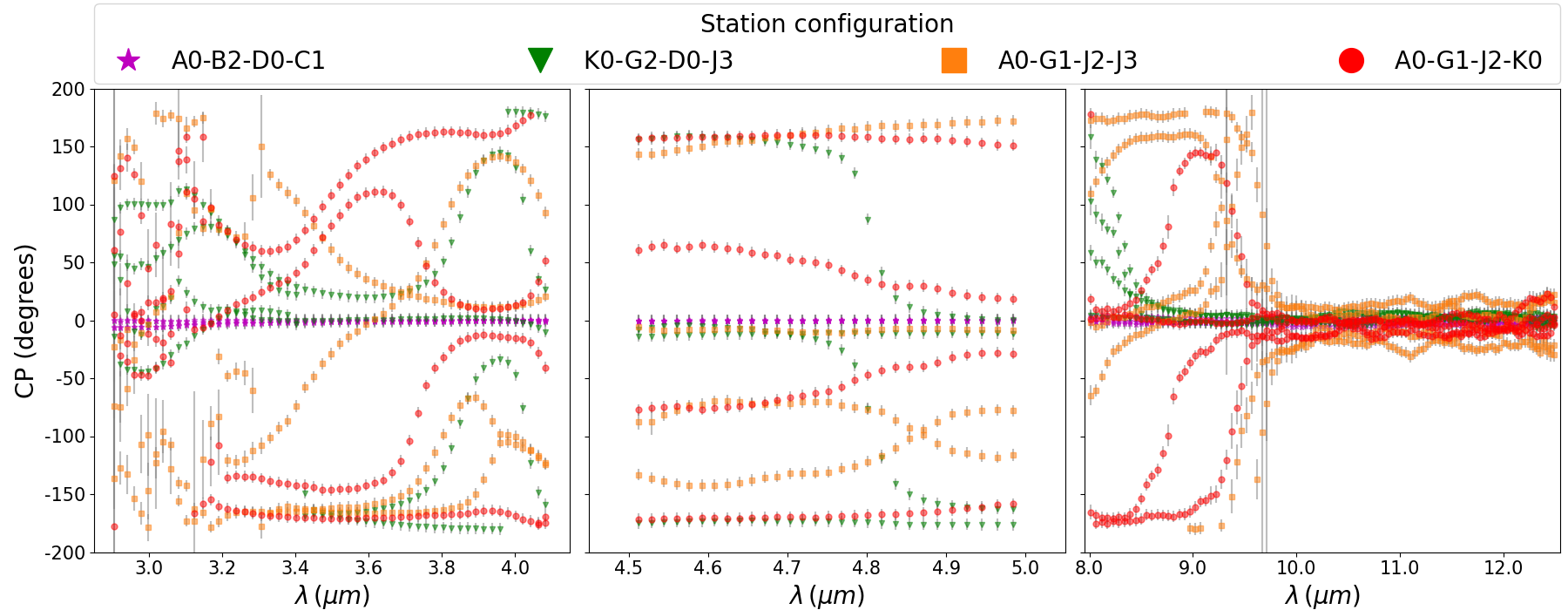}
    \caption{Comparison of the closure phases from the MATISSE observations with wavelength. The markers colour represent the AT configuration corresponding to each curve.}
    \label{fig:CPH}
\end{figure*}

\begin{figure*}[h!]
    \centering
    \includegraphics[width=\linewidth]{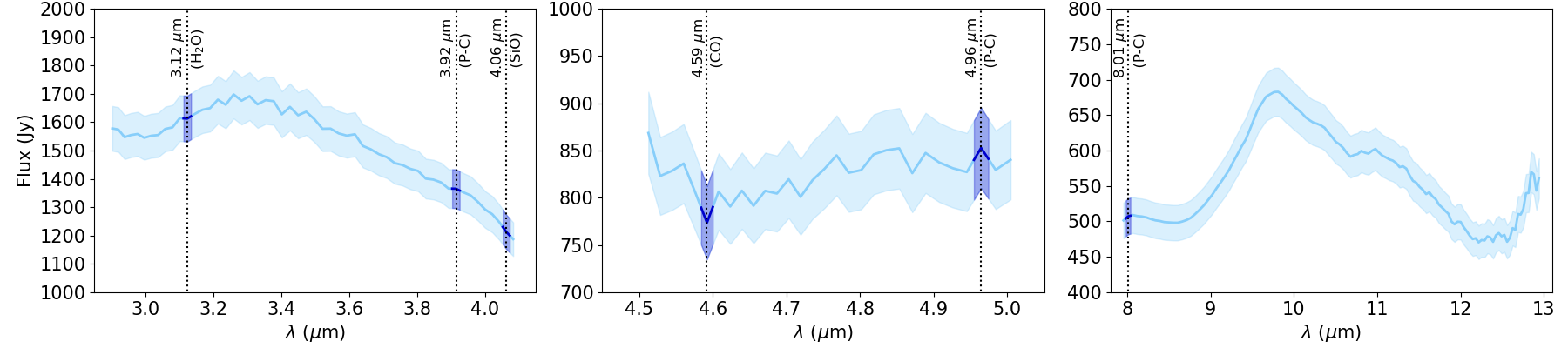}
    \caption{MATISSE calibrated spectrum of the L-, M- and N-bands (from left to right). The wavelength ranges used for the image reconstruction are shown in dark blue.}
    \label{fig:MATISSE_spectra}
\end{figure*}

Figure~\ref{fig:VIS2} shows the squared visibilities (V$^{2}$) in logarithmic scale against the wavelength. Figure~\ref{fig:CPH} shows the closure phases (CP) against the wavelength. It is worth noting that the CP curves become too noisy below $\sim$3.1\,\mum\ and almost flat beyond $\sim$9\,\mum, in particular for the most compact configurations. Figure~\ref{fig:MATISSE_spectra} shows the MATISSE absolutely-photometrically calibrated spectra in the three different bands.

\section{MATISSE wavelength-resolve images}\label{images}

\begin{figure*}
    \centering
    \includegraphics[width=\linewidth]{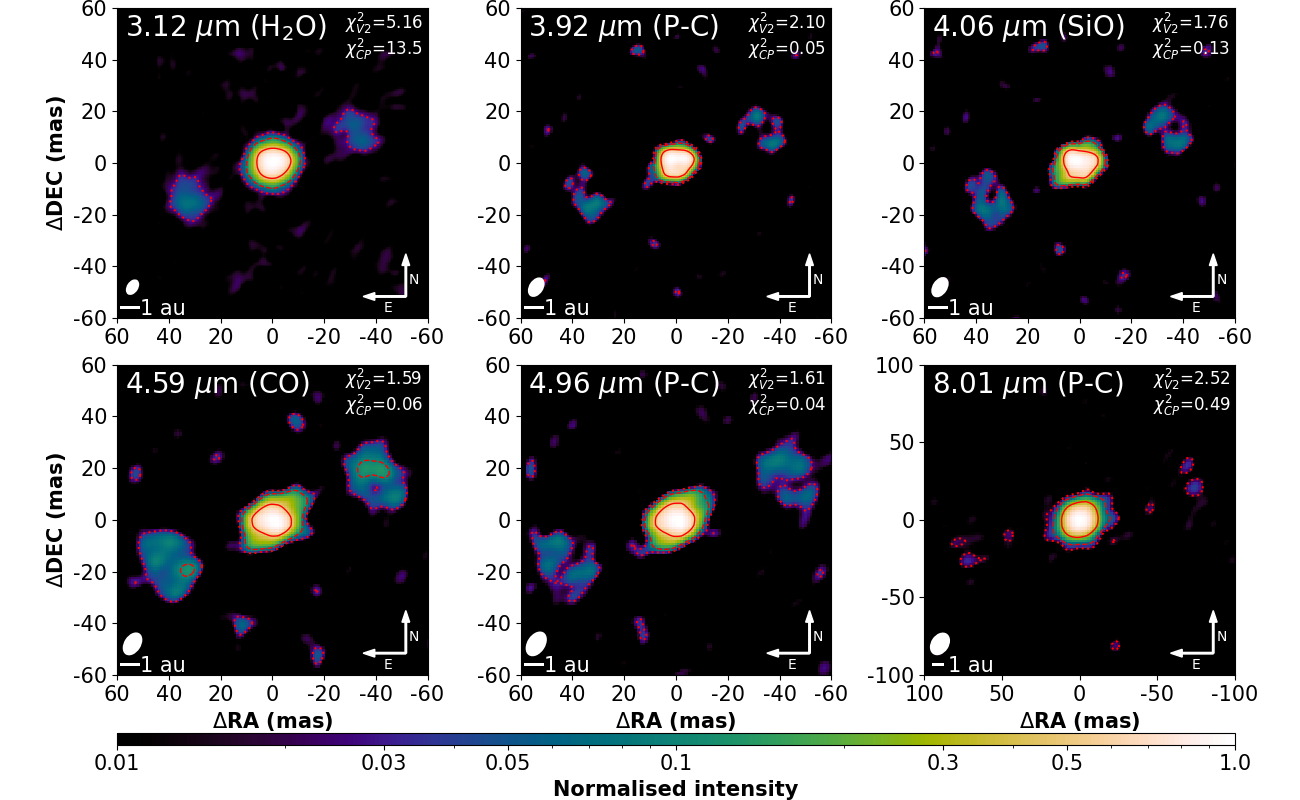}
    \caption{Set of MATISSE reconstructed images of EP\,Aqr, in the pseudo-continuum (P-C) and molecular bands. In each panel, the wavelength, label (P-C or molecular band), and $\chi^{2}$ (for both the V$^{2}$ and the CP) are shown at the top, whereas the beam size and linear scale (1 au) are shown at the bottom-left as white ellipses and lines, respectively. Red contours (solid, dashed, and dotted) represent the 3, 10, and 50 rms levels to highlight the changes of the central disc and the blobs between the images. Note the different fields of view between the images in the LM-bands (3.12, 3.92, 4.06, 4.59, and 4.96 \mum) and the image in the N-band (8.01 \mum).}
        \label{MATISSE_images}
\end{figure*}

\begin{figure*}[h!]
     \centering
     \begin{subfigure}[b]{0.475\linewidth}
         \centering
         \includegraphics[width=\linewidth]{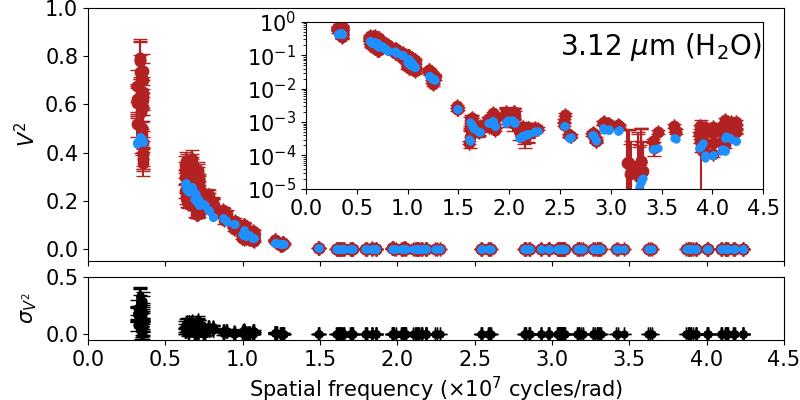}      
     \end{subfigure}
     \begin{subfigure}[b]{0.475\linewidth}
         \centering
         \includegraphics[width=\linewidth]{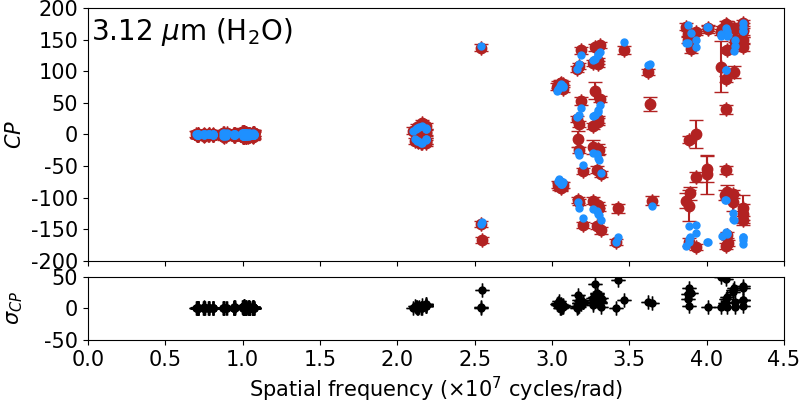}
     \end{subfigure}
     \begin{subfigure}[b]{0.475\linewidth}
         \centering
         \includegraphics[width=\linewidth]{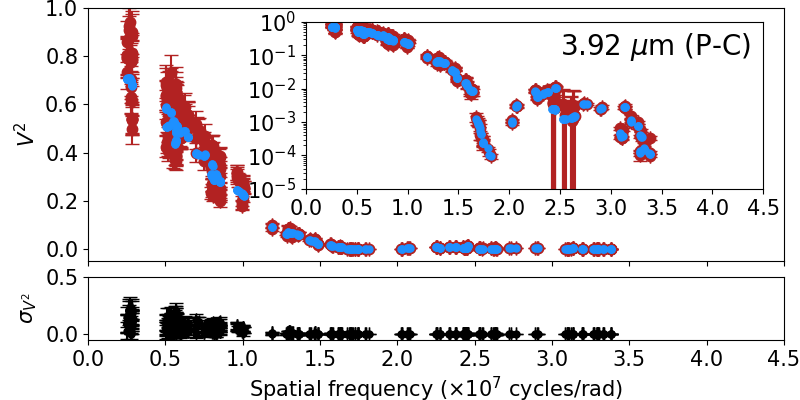}      
     \end{subfigure}
     \begin{subfigure}[b]{0.475\linewidth}
         \centering
         \includegraphics[width=\linewidth]{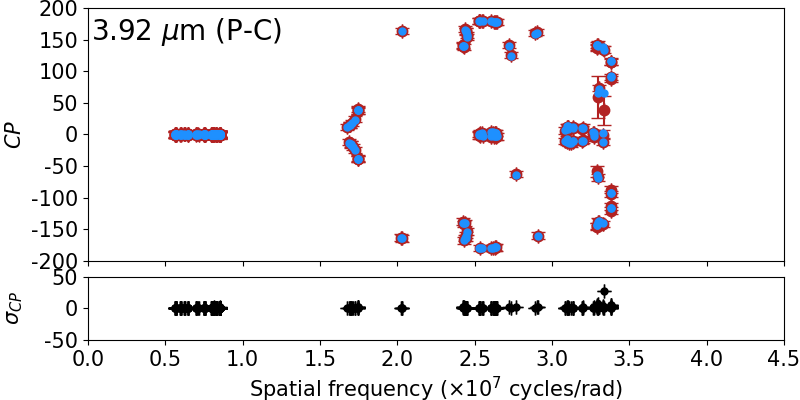}
     \end{subfigure}
     \caption{Representation of the MATISSE observed (red) and MiRA reconstructed (blue) observables for the different images presented in Fig.~\ref{MATISSE_images}. From top to bottom: (i) 3.12 \mum\,(H$_{2}$O), (i) 3.92 \mum\,(L-band P-C), (i) 4.06 \mum\,(SiO), (i) 4.59 \mum\,(CO), (i) 4.96 \mum\,(M-band P-C), and (i) 8.01 \mum\,(N-band P-C). {\it Left}: Squared visibilities. The inset panels show the same curve on a logarithmic scale. {\it Right}: Closure phases. All panels are accompanied by a plot with the residual of the fit ($\sigma_{\mathrm{V}^{2}}$ and $\sigma_{\mathrm{CP}}$, respectively).}
     \label{fig:observales}
\end{figure*}

Figure~\ref{MATISSE_images} shows the set of six MATISSE reconstructed images at specific wavelengths used in this study. Figure~\ref{fig:observales} shows the comparison between the observed V$^{2}$ and CP curves with those corresponding to the reconstructed images against the spatial frequency for the different wavelengths of the images presented in Fig.~\ref{MATISSE_images}. It can be directly appreciated that the first minimum in the V$^{2}$ curves shifts to lower spatial frequencies as the wavelength increases, which is dominated by the increase in the angular size of the central source, except for the 3.12\,\mum\ image that shows the opposite trend. 

\begin{figure*}[h!]
     \centering
     \begin{subfigure}[b]{0.475\linewidth}
         \centering
         \includegraphics[width=\linewidth]{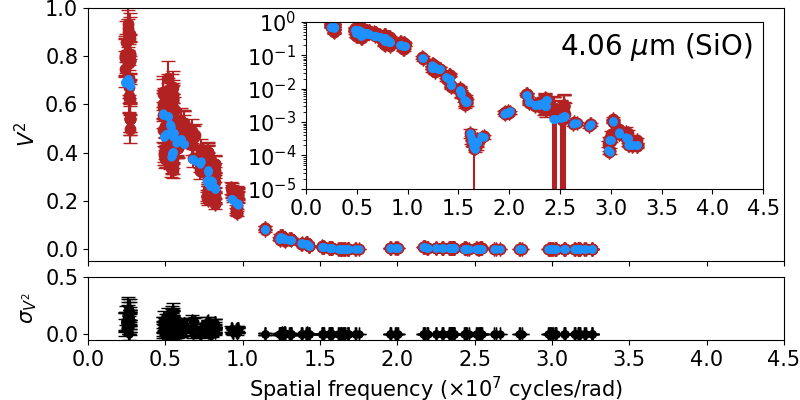}      
     \end{subfigure}
     \begin{subfigure}[b]{0.475\linewidth}
         \centering
         \includegraphics[width=\linewidth]{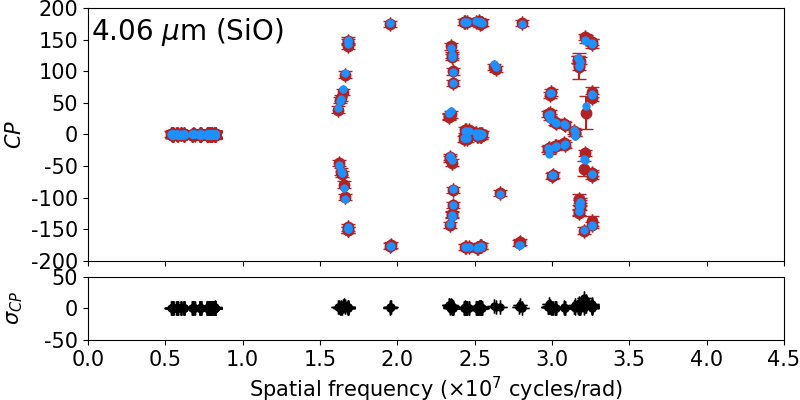}
     \end{subfigure}
     \begin{subfigure}[b]{0.475\linewidth}
         \centering
         \includegraphics[width=\linewidth]{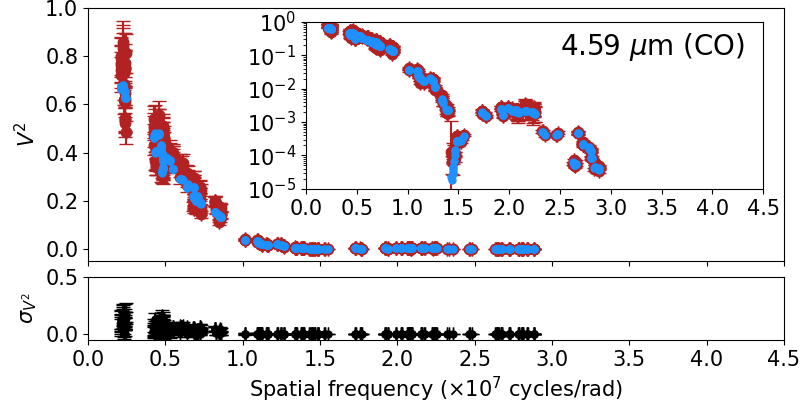}      
     \end{subfigure}
     \begin{subfigure}[b]{0.475\linewidth}
         \centering
         \includegraphics[width=\linewidth]{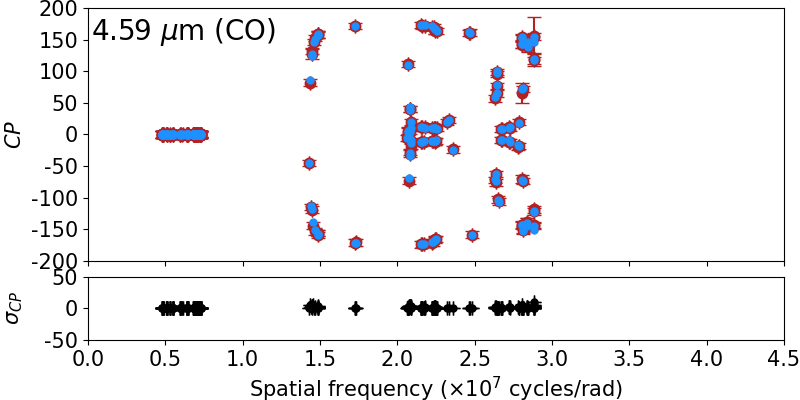}
     \end{subfigure}
     \begin{subfigure}[b]{0.475\linewidth}
         \centering
         \includegraphics[width=\linewidth]{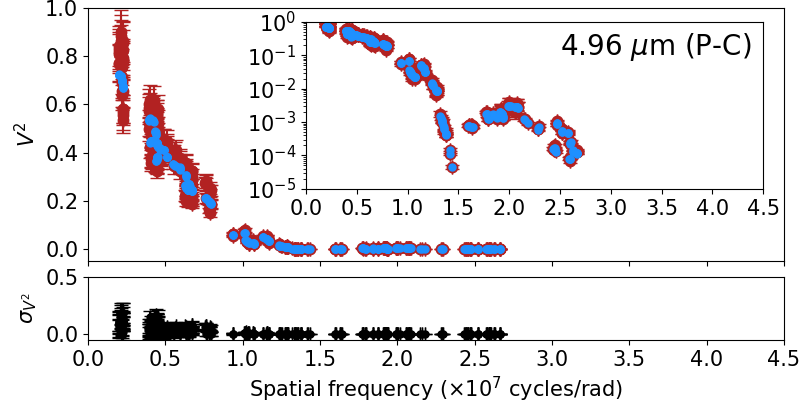}      
     \end{subfigure}
     \begin{subfigure}[b]{0.475\linewidth}
         \centering
         \includegraphics[width=\linewidth]{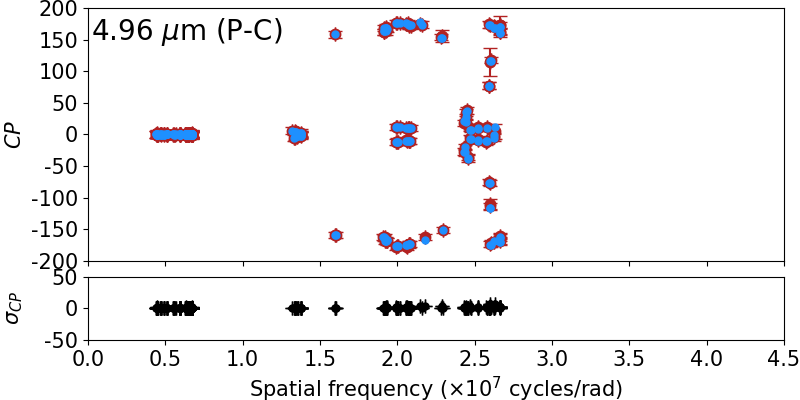}
     \end{subfigure}
     \begin{subfigure}[b]{0.475\linewidth}
         \centering
         \includegraphics[width=\linewidth]{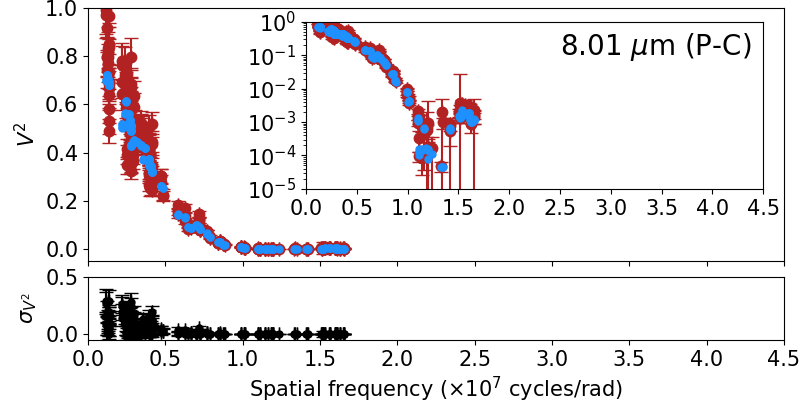}      
     \end{subfigure}
     \begin{subfigure}[b]{0.475\linewidth}
         \centering
         \includegraphics[width=\linewidth]{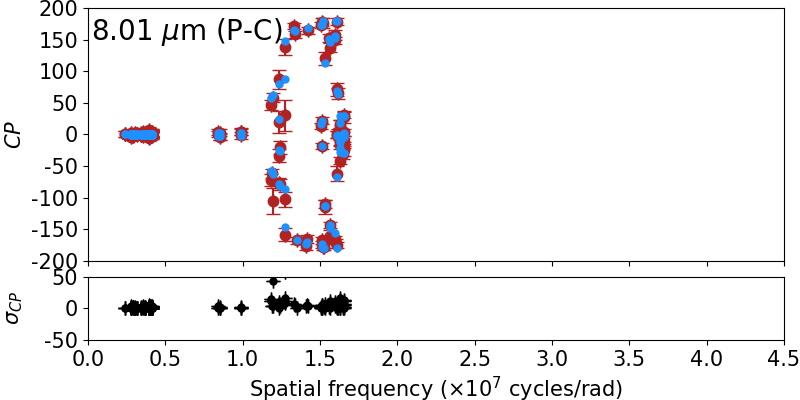}
     \end{subfigure}
     \caption*{Fig.~\ref{fig:observales} continued.}
     \label{fig:observales2}
\end{figure*}

\section{Geometric modelling of the visibilities}\label{UV_fitting}

Given the sparse ($u$,$v$)-plane coverage, in particular in the direction of the blobs, we performed some tests to check whether they are real structures or artifacts from the image reconstruction. We performed geometric modelling in the ($u$,$v$)-plane with the \texttt{oimodeler}\footnote{\url{https://github.com/oimodeler/oimodeler}} tool \citep{Meilland_2024}. We used the 4.59\,\mum\ observables as a reference for the modelling. We found that the complex structure in the data makes it impossible to achieve a good solution using simple models (e.g. two elongated discs). Therefore, we used a three-component model composed of: (i) a uniform disc, (ii) an elongated disc, and (iii) an elongated ring. 

\begin{figure*}[h!]
    \centering
    \includegraphics[width=0.7\linewidth]{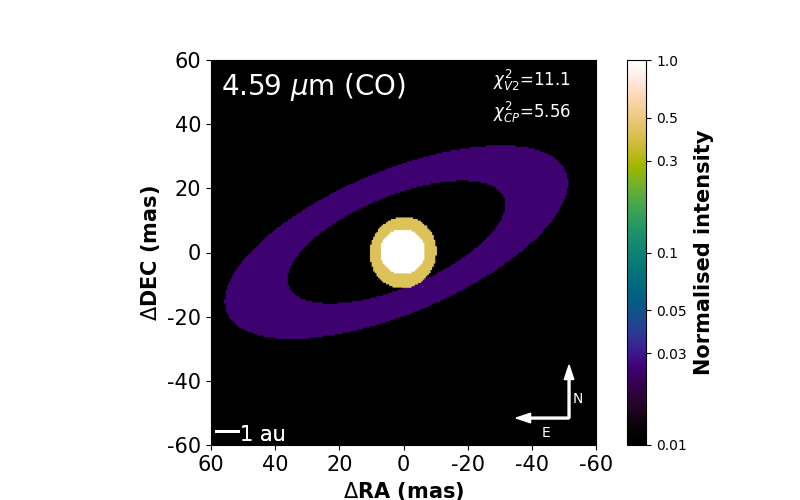}
    \caption{Representation of the best-fit geometric model at the 4.59\,\mum. The wavelength, label, and $\chi^{2}$ (for both the V$^{2}$ and the CP) are shown at the top, whereas the linear scale (1\,au) is shown at the bottom-left as a white line.}
    \label{fig:model_image}
\end{figure*}

Figure~\ref{fig:model_image} shows the resulting best-fit solution. This model illustrates that the two first components of the model represent the central source, whereas an extensive ring represents the blobs found in the reconstructed images. We highlight that the inner diameter of the elongated ring converges toward a position close to that of the blobs in that image ($\sim$30 mas) rather than lower values. Table~\ref{tab:model_parameters} shows the parameters of the best-fit solution.

\begin{figure*}[h!]
     \centering
     \begin{subfigure}[b]{0.475\linewidth}
         \centering
         \includegraphics[width=\linewidth]{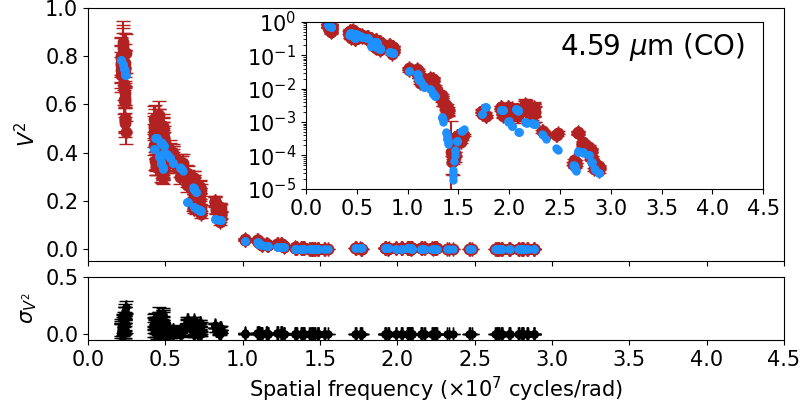}      
     \end{subfigure}
     \begin{subfigure}[b]{0.475\linewidth}
         \centering
         \includegraphics[width=\linewidth]{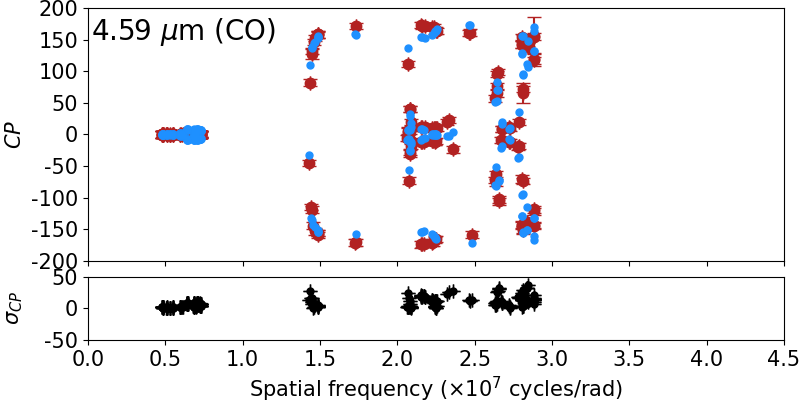}
     \end{subfigure}
     \caption{Same as Fig.~\ref{fig:observales}, but for the best-fit geometric model.}
     \label{fig:model_observales}
\end{figure*}

Figure~\ref{fig:model_observales} shows the comparison between the modelled observables and those obtained with MATISSE, this model can correctly fit the observables, in particular at medium and high spatial frequencies. Therefore, this structure is necessary to fit the observations, or at least it is consistent with them. 

From this geometric model, we derive an elongation (i.e. the ratio between the major and the minor axes) of the disk-like structure, represented by the elongated ring, of $\sim$2.56. Under the assumption that the structure is intrinsically circular and that its apparent elongation results from inclination effects, we estimate an inclination angle over the sky of $\sim$70\degree (or $\sim$110\degree, since the projected elongation does not provide information on the direction of the inclination axis).

Finally, we also found that when using different subsamples of the dataset, which change the predominant direction of the synthesised beam and, therefore, any potential artifacts, the location of the blobs does not suffer significant changes. Therefore, the presence of the blobs in the reconstructed images seem to be a robust component and not a spurious artifact produced by the sparse ($u$,$v$)-plane coverage.

\begin{table*}[h!]

\renewcommand{\arraystretch}{1.1}
\small
\centering

\caption{Parameters of the best-fit geometric model obtained at 4.59 \mum.} 
\label{tab:model_parameters}

\begin{adjustbox}{max width=\textwidth}
\begin{threeparttable}[b]

\begin{tabular}{c c c c c c c c c c c c c c c}
\hline\hline 
$f_{1}$ & $d_{1}$ & $f_{2}$ & $x_{2}$ & $y_{2}$ & $e_{2}$ & PA$_{2}$ & $d_{2}$ & $f_{3}$ & $x_{3}$ & $y_{3}$ & $e_{3}$ & PA$_{3}$ & $d^{inn}_{3}$ & $d^{out}_{3}$ \\  

  & (mas) &  & (mas) & (mas) &  & (\degree) & (mas) &   & (mas) & (mas) &  & (\degree) & (mas) & (mas) \\  
\hline 
0.29 & 13.99 & 0.51 & -0.15 & -0.23 & 1.08 & 164.62 & 22.45 & 0.20 & 1.97 & 2.99 & 2.56 & 112.64 & 72.26 & 114.52\\

\hline
\end{tabular} 
\begin{tablenotes}
\item \textbf{Notes.} Definition of the model parameters. $f_{1}$: Relative flux of the central uniform disc, $d_{1}$: Diameter of the central uniform disc, $f_{2}$: Relative flux of the elongated disc, $x_{2}$: $\Delta$RA of the elongated disc, $y_{2}$: $\Delta$DEC of the elongated disc, $e_{2}$: Elongation of the elongated disc, PA$_{2}$: Position angle of the elongated disc, $d_{2}$: Diameter of the elongated disc, $f_{3}$: Relative flux of the elongated ring, $x_{3}$: $\Delta$RA of the elongated ring, $y_{3}$: $\Delta$DEC of the elongated ring, $e_{3}$: Elongation of the elongated ring, PA$_{3}$: Position angle of the elongated ring, $d^{inn}_{3}$: Inner diameter of the elongated disc, $d^{out}_{3}$: Outer diameter of the elongated disc.
\end{tablenotes}

\end{threeparttable}
\end{adjustbox}

\renewcommand{\arraystretch}{1.0}

\end{table*}

\section{ALMA continuum emission}\label{ALMA_continuum}

We include the analysis of the ALMA continuum observations at $\sim$230\,GHz. To perform the comparison between MATISSE images and ALMA maps, we centred the emission peaks of the central sources in the MATISSE and ALMA continuum images.
 
\begin{figure*}[h!]
     \centering
     \begin{subfigure}[b]{0.48\linewidth}
         \centering
         \includegraphics[width=\linewidth]{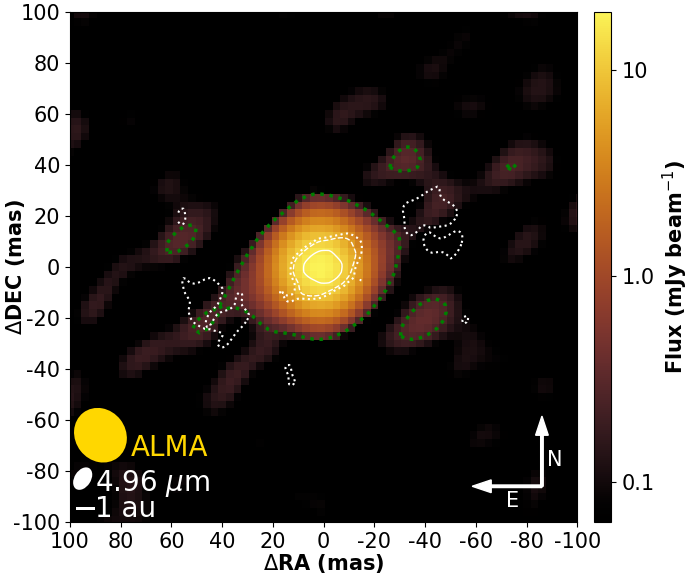}
     \end{subfigure}
     \begin{subfigure}[b]{0.48\linewidth}
         \centering
         \includegraphics[width=\linewidth]{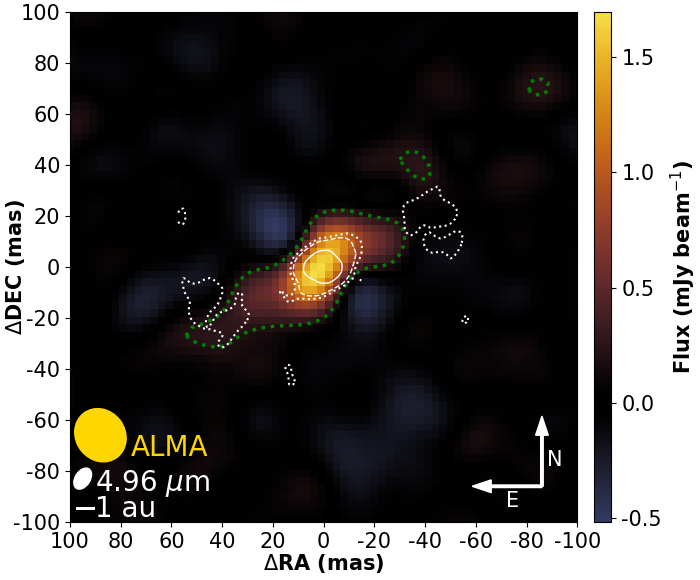}
     \end{subfigure}
        \caption{\textit{Left:} ALMA 230\,GHz self-calibrated continuum map. \textit{Right:} ALMA 230\,GHz self-calibrated continuum map after subtracting a circular homogeneous disc. Green contours represent the 3 rms levels. White contours (solid, dashed, and dotted) represent the 3, 10, and 50 rms levels of the 4.96\,\mum\ image. Beam sizes and linear scale (1\,au) are shown at the bottom-left as coloured ellipses and a white line.}
        \label{fig:MATISSE+cont}
\end{figure*}

Fig.~\ref{fig:MATISSE+cont} (left) shows the continuum emission at 230\,GHz. Although the central source remains relatively compact at the angular resolution of these observations, it exhibits a clear elongation along the same position angle as the central source identified in the MATISSE images that is not caused by the synthesised beam. We also note the presence of weak extended emission, although its intensity is comparable to the noise level. With the aim of better showing the elongation of the central source without introducing a bias in its orientation, we fitted and subtracted a circular homogeneous disc model using the \texttt{Gildas} \texttt{UVFIT} procedure to directly fit in the Fourier space. We found the best-fit disc with a diameter of 14.87$\pm$0.15\,mas (i.e. 1.92$\pm$0.02\,au) and a flux of 21.03$\pm$0.06\,mJy. We noted that in the case of an elongated disc model, the position angle was degenerated and could not be accurately constrained. Finally, we constrained an upper limit in the integrated flux of the blobs of $<$0.17 mJy (i.e. $<$0.8\% with respect to the central source) by integrating the flux in apertures with the size of the ALMA beam in the position of the blobs, whereas it is $\sim$18\% and $\sim$4\% in the 4.59\,\mum\ and 8.01\,\mum\ images, respectively. This spectral drop in the flux of blobs implies a spectral index $\nu$$\gtrsim$2.2.

Fig.~\ref{fig:MATISSE+cont} (right) shows the residual image after subtracting the circular homogeneous disc. The residual shows a clearly elongated structure with positive values in the direction of the blobs, whereas it has negative values in the perpendicular direction. This result shows the presence of a faint and elongated secondary component that is clearly aligned with the blobs. Nevertheless, it cannot be determined whether this component is part of an inhomogeneous stellar surface or it is related with the blobs due to the angular resolution of the available observations.

\end{appendix}

\end{document}